%% file: 0_main.tex
\documentclass[sigconf]{acmart}

\AtBeginDocument{%
  \providecommand\BibTeX{{%
    \normalfont B\kern-0.5em{\scshape i\kern-0.25em b}\kern-0.8em\TeX}}}

\AtBeginDocument{%
 \providecommand\BibTeX{{%
    \normalfont B\kern-0.5em{\scshape i\kern-0.25em b}\kern-0.8em\TeX}}}
    
\usepackage{enumitem}
\usepackage{multirow}
\usepackage{makecell}
\usepackage{amsthm}
\usepackage{amsfonts}

\setlist[itemize]{leftmargin=*}
\setcopyright{acmcopyright}
\begin{document}
\fancyhead{}

\copyrightyear{2021}
\acmYear{2021}
\setcopyright{acmcopyright}
\acmConference[ICMR '21] {Proceedings of the 2021 International Conference on Multimedia Retrieval}{August 21--24, 2021}{Taipei, Taiwan.}
\acmBooktitle{Proceedings of the 2021 International Conference on Multimedia Retrieval (ICMR '21), August 21--24, 2021, Taipei, Taiwan}
\acmPrice{15.00}
\acmISBN{978-1-4503-8463-6/21/08}
\acmDOI{10.1145/3460426.3463638}

\title{Leveraging Two Types of Global Graph for Sequential Fashion Recommendation}

 \author{Yujuan Ding}
 \affiliation{
     \institution{Shenzhen University}
 }
 \email{dingyujuan385@gmail.com}
 
\author{Yunshan Ma}
 \affiliation{
     \institution{National University of Singapore}
 }
 \email{yunshan.ma@u.nus.edu}

 \author{Wai Keung Wong}
 \authornote{Corresponding author.}
 \affiliation{
    \institution{The Hong Kong Polytechnic University}
 }
 \email{calvin.wong@polyu.edu.hk}

 \author{Tat-Seng Chua}
 \affiliation{
     \institution{National University of Singapore}
 }
 \email{dcscts@nus.edu.sg}

\begin{abstract}
Sequential fashion recommendation is of great significance in online fashion shopping, which accounts for an increasing portion of either fashion retailing or online e-commerce. The key to building an effective sequential fashion recommendation model lies in capturing two types of patterns: the personal fashion preference of users and the transitional relationships between adjacent items. The two types of patterns are usually related to user-item interaction and item-item transition modeling respectively. However, due to the large sets of users and items as well as the sparse historical interactions, it is difficult to train an effective and efficient sequential fashion recommendation model. To tackle these problems, we propose to leverage two types of global graph, \textit{i.e.}, the user-item interaction graph and item-item transition graph, to obtain enhanced user and item representations by incorporating higher-order connections over the graphs. In addition, we adopt the graph kernel of LightGCN~\cite{he2020lightgcn} for the information propagation in both graphs and propose a new design for item-item transition graph. Extensive experiments on two established sequential fashion recommendation datasets validate the effectiveness and efficiency of our approach.
\end{abstract} 


\maketitle
\pagestyle{plain}

\input{1_introduction}

\input{2_related_work}
\input{3_approach}
\input{4_experiments}
\input{5_conclusion}

\input{6_acknowledgement}
\newpage
\bibliographystyle{ACM-Reference-Format}
\bibliography{0_main}

\end{document}

%% file: 1_introduction.tex
\section{Introduction}
\label{sec:intro}
Recommendation system has become an essential feature in many online platforms, especially for those that connect users and items. For online fashion shopping, fashion recommender system can improve users' shopping experience and retailers' sales volume by routing users to their preferred fashion items. Therefore, it is of great value to develop powerful fashion recommendation models, and sequential fashion recommendation is one of the key techniques to achieve this goal. 

\begin{figure}
    \centering
    \includegraphics[width = 0.9\linewidth]{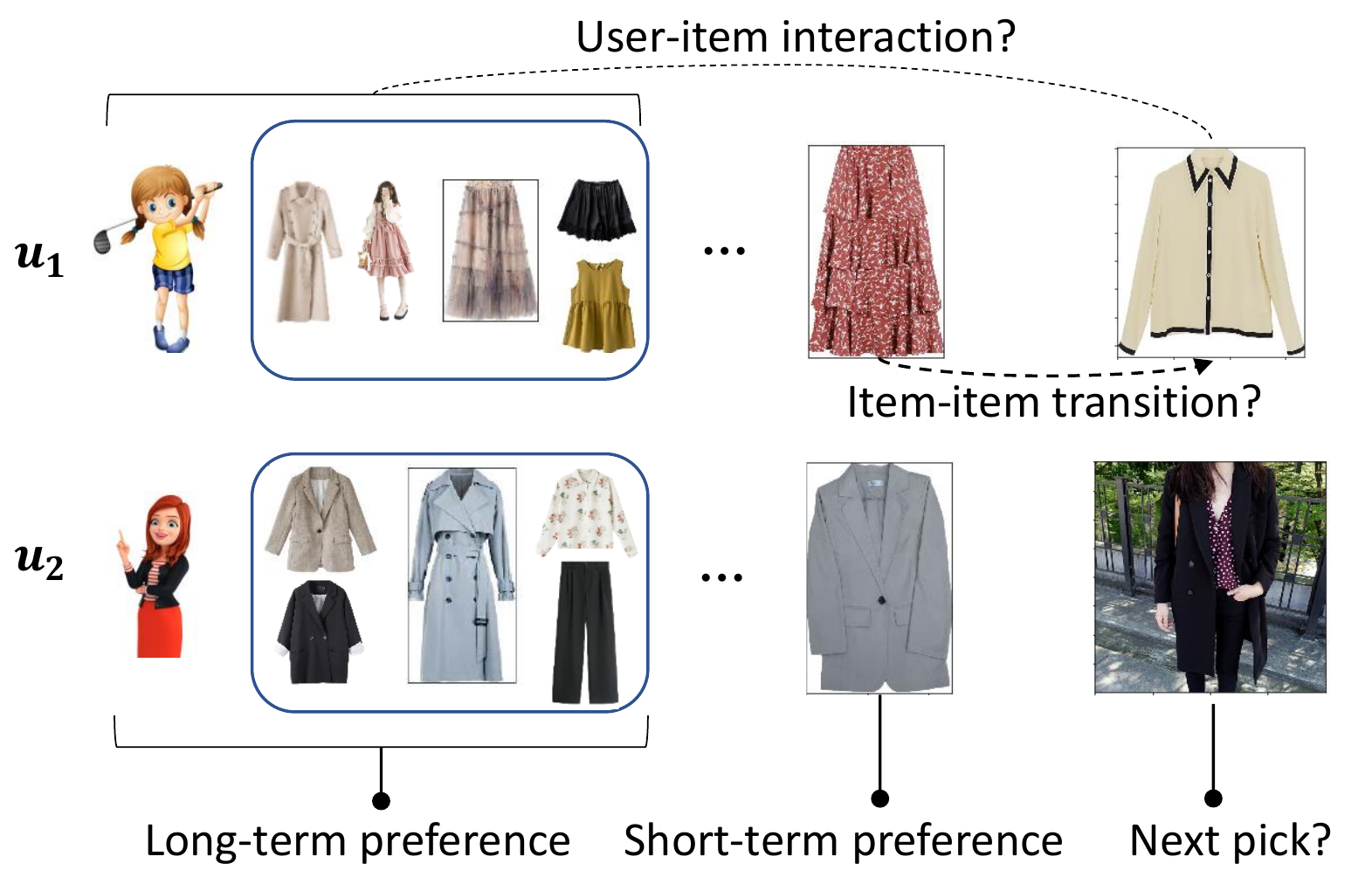}
    \caption{An illustration of the sequential fashion recommendation task, which aims to predict the user's next interaction by exploring both the long-term (u-i interaction) and short-term (i-i transition) fashion preference.}
    \label{fig:task}
    \vspace{-10pt}
\end{figure}

The key to build an effective sequential recommender system for fashion lies in capturing both the user's personalized fashion preference patterns and item's transitional patterns. User's actions (\textit{e.g.}, clicking or buying) on the online fashion shopping platforms naturally form the chronological sequences. 
To predict the next interacted item of a user, we should predict not only the user-item (abbreviated to u-i) interaction probability (related to long-term user preference) but also the item-item (abbreviated to i-i) transition probability between his/her previous choice(s) and the next one. As shown in Figure 1, $u_1$ is a young girl who likes skirts and dresses, $u_2$ is a middle-aged business woman who prefers business-style clothes. Such personal preference is long-term and static, which can be explored from the overall historical behavior of users. On top of that, users' short-term interest also affects their choice at specific time point. 
We can see from the example that just before the particular time point of the recommendation, $u_1$ bought a red skirt. Her next action could be highly related to the previous one, for example, she might want to find an item to match with the skirt, which makes the shirt a proper recommendation. Similarly, the two items interacted by $u_2$ are also related with each other, where she would like a black suit to substitute the previous grey one. To sum up, both the u-i interaction and i-i transition patterns are prevalent in users' online shopping experience, and how to properly model both of them within one unified model is a key research problem.

To properly model both the u-i interaction and i-i transition, there are several challenges. First, the large set of items and sparse historical interactions result in severe data sparsity in fashion domain. As shown by the real-world fashion datasets in Table~\ref{tab:dataset}, the number of items is over hundreds of thousands, while the number of interactions per item is extremely small. As a consequence, traditional matrix factorization-based methods, such as MF~\cite{rendle2012bpr} and FPMC~\cite{rendle2010factorizing}, are unable to effectively model the u-i interaction and i-i transition patterns. Even worse, for methods that require a sequence of interacted items as input, such as GRU4REC~\cite{hidasi2015session} and Caser~\cite{tang2018personalized}, the number of available training samples will be further reduced. Therefore, it is difficult for those models to capture the inherent patterns on such sparse datasets. Second, most existing approaches that models i-i interactions as graphs are inefficient due to either inappropriate construction of the graph or complicated graph kernels. 
For example, SR-GNN~\cite{wu2019session} only models the session-level i-i transition graph while fails to take into account the effect between items from different sessions. Moreover, the graph kernels that have been widely applied in many existing methods, such as Gated Graph Neural Network (GGNN)~\cite{li2015gated,wu2019session} or Graph Attention Network (GAT)~\cite{velivckovic2017graph,qiu2019rethinking}, are complicated, resulting in more computational costs. In summary, how to tackle the problems of data sparsity and design an effective yet efficient i-i transition graph are the most critical considerations when designing the models.

In this paper, to deal with the above mentioned challenges, we propose the \textbf{D}ual-\textbf{G}raph \textbf{S}equential \textbf{R}ecommender (DGSR) method. 
First, to counter the data sparsity problem, we leverage two types of graph based on the global u-i interactions and i-i transitions. By performing information propagation over the two global graphs, both the user and item representations will be enhanced by the global u-i collaborative filtering (CF) signals and the i-i transition contexts. In addition, higher-order propagation extends the connections in the current graph and further relieves the data sparsity problem. Second, we propose a new design which formulates the i-i transition as a bipartite graph. The graph assigns each item with two nodes -- one for the situation when the item serves as an anchor and another for the situation when the item serves as a target. We drop the session-level graph and purely utilize the global graph for efficiency consideration. We adopt the LightGCN~\cite{he2020lightgcn} as the graph kernel to further reduces the computational costs. LightGCN revises the typical Graph Convolutional Network (GCN)~\cite{kipf2016gcnsemi} model by removing the feature transformation layer and the non-linear activation layer, which improves the performance and also reduce the computational cost.
The main contributions of this work are summarized as follows: 
\begin{itemize}
\item We leverage two types of global graph to relieve the data sparsity problem and enhance the user and item representations. To the best of our knowledge~\cite{wu2020graph}, this is the first work to leverage both u-i and i-i global graphs for sequential fashion recommendation.
\item We propose a new design for the i-i transition graph construction and adopt LightGCN as the kernel in both the u-i and i-i graphs, which lead to great improvement in performance and efficiency.
\item Extensive experiments on two established fashion sequential recommendation datasets (\textit{i.e.}, Taobao iFashion and Amazon Fashion) demonstrate the effectiveness of our proposed method.
\end{itemize}

\begin{table}[!t]
\caption{Dataset Statistics}
\vspace{-0.15in}
\begin{center}
\setlength{\tabcolsep}{2.3mm}{\begin{tabular}{{ccc|cc}}
\hline
Dataset & \multicolumn{2}{c|}{iFashion-SR} &\multicolumn{2}{c}{Amazon-Fashion}\\
Min Seq Len & Seven  &Ten & Seven &Ten \\ 
\hline
\#User & 36,752 & 36,797 &48,427 &19,362 \\
\#Item & 458,642 & 460,596 &137,650 &106,256 \\
\#Actions &1,624,643 &1,639,006 &509,352 &284,385 \\ 
\#Train sample &1,474,640 &1,489,000 &364,071 &226,299 \\ 
\#Test sample &50,001 &50,002 &48,427 &19,362 \\
\#Valid sample &50,001 &50,002 &48,427 &19,362 \\

avg. \#act/user &44.2 &44.54 &10.52 &14.69 \\ 
avg. \#act/item &3.54 &3.56 &3.70 &2.68 \\ 

\hline
\end{tabular}}
\vspace{-0.15in}
\label{tab:dataset}
\end{center}
\end{table}

%% file: 2_related_work.tex
\section{Related Work}
\label{sec:relatedwork}
\subsection{Sequential and Fashion Recommendation}
Personalized fashion recommendation aims to model the personalized fashion preference of users. Most previous works~\cite{he2016sherlock,he2016ups, yu2018aesthetic, vbpr, kang2017visually, liu2017deepstyle} focus on emphasizing visual representation of fashion items in building the recommender system as visual information is considered much more important in fashion domain than in other domains. Despite the effectiveness of incorporating visual feature, existing works widely overlook the unique user behaviors in fashion domain and fail to explore more effective behavior patterns from the interaction history of users. The sparsity issue is also not properly addressed in previous works. However, the issue is particularly significant in fashion domain as the item set is way larger compared to other domains such as books or music~\cite{hu2015collaborative}.

In literature, matrix factorization (MF)~\cite{rendle2012bpr} is one of the most simple and effective non-sequential recommendation methods. However, MF, as well as other non-sequential methods~\cite{wang2019ngcf,he2020lightgcn,NCF}, only models the u-i interactions but is not able to take i-i transition into account. In comparison, session-based methods usually solely explore the item transition patterns for the recommendation. In recent years, Deep Neural Networks (DNNs) have been widely applied in session-based recommendation for its capability of modeling sequential data. Various types of DNNs, e.g., Recurrent Neural Networks (RNNs), Convolutional Neural Networks (CNNs), have been applied in developing advanced session-based recommenders~\cite{hidasi2015session, li2017neural, quadrana2017personalizing, tang2018personalized}. In the recent years, GNNs have drawn special attention and some GNN-based methods have achieved state-of-the-art performance in session-based recommendation tasks~\cite{wu2020graph}.

Compared with above two types of methods, sequential recommendation methods model the user's general preference and also analyze the latest action(s) of him~\cite{fang2019deep,quadrana2018sequence}. Markov Chain (MC) is an effective tool to model the action sequence of users and infer the next one. Combined with Matrix Factorization (MF), Rendel et al.~\cite{rendle2010factorizing} propose the Factorized Personalized Markov Chain (FPMC) to capture both the sequential patterns and the long-term user preference. Despite of its simplicity and efficiency ~\cite{feng2015personalized,wang2015learning,he2016fusing}, it only applies the first-order MC and fail to model any high-order user-item or item-item dependency or connectivity, which still limits the overall recommendation performance. 

\begin{figure*}
    \centering
    \includegraphics[width = 0.9\linewidth]{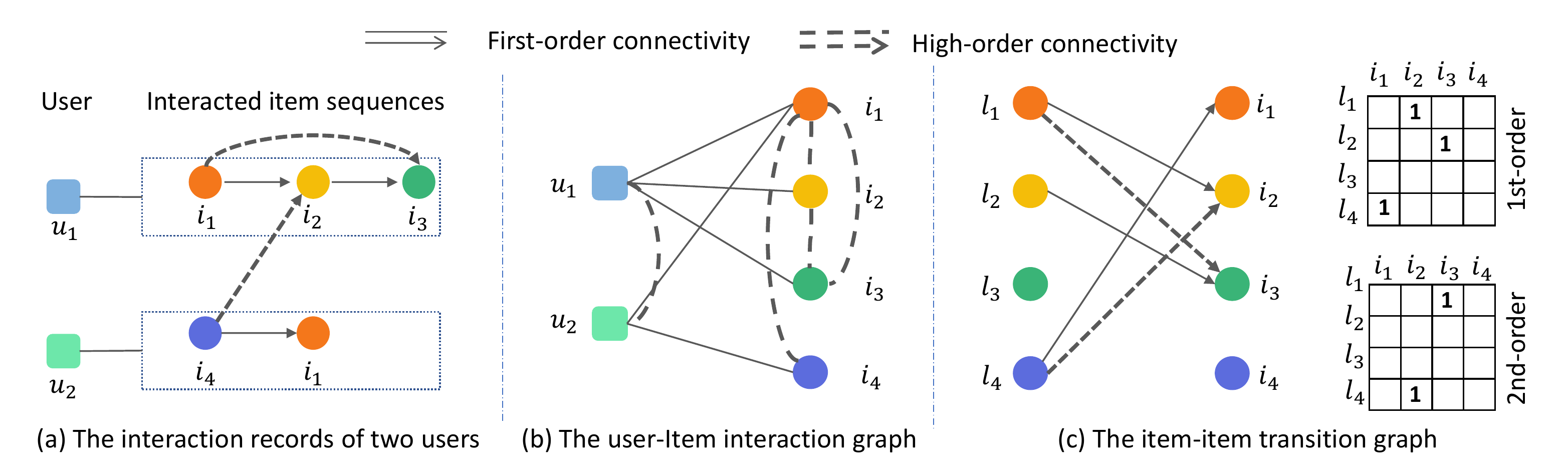}
    \vspace{-0.1in}
    \caption{An illustration of the user-item interaction graph and item-item transition graph in sequential recommendation.}
    \vspace{-0.1in}
    \label{fig:graph}
\end{figure*}

\subsection{Graph Neural Networks for Recommendation}
With the rapid development and tremendous success of GNNs in many application domains~\cite{scarselli2008graph,kipf2016gcnsemi,ding2021leveraging}, recommendation approaches based on GNNs~\cite{wu2020graph,guo2020survey} have achieved state-of-the-art performance in various sub-tasks, such as implicit feedback-based general recommendation and session-based recommendation~\cite{he2020lightgcn,wang2020global}. Graph is a natural data structure for most of data in recommender systems. For example, the user and item interactive relationship can be modeled with a bipartite graph. Another main contribution of GNN in recommender system is its ability of capturing the high-order connectivity, which can help inject the CF signal into the embedding learning process, thereby achieving significant improvements on performance~\cite{wang2019ngcf, he2020lightgcn}. Wang et al.~\cite{wang2019ngcf, he2020lightgcn} proposed the Neural Graph Collaborative Filtering (NGCF) method for the implicit feedback-based recommendation. NGCF builds the user-item interaction graph first and then conducts multi-layer embedding propagation on the graph to refine the embeddings of users (or items). It has achieved very competitive performance over all CF-based recommendation methods, as well as its variant LightGCN~\cite{he2020lightgcn}. 

GNNs have also been applied in session-based recommendation to capture the transition patterns in sessions. The majority of existing methods build the graphs based on the sequence of items within the same session and then apply GNNs to capture transitions among items, which are claimed to be complex and difficult to be captured by previous conventional sequential methods~\cite{wu2019session}. Methods with such a framework include SR-GNN~\cite{wu2019session}, FGNN~\cite{qiu2019rethinking}, GC-SAN~\cite{xu2019graph}, A-PGNN~\cite{wu2019personalized}. In summary, different efforts have been made in modeling the high-order connectivity of either user-item interaction or item-item transition. However, so far no work has tried to modeling both yet. Since the sparsity issue is particularly severe in sequential fashion recommendation, leveraging more contextual information by high-order connectivity, for both u-i interaction and i-i transition, would be significantly valuable.

%% file: 3_approach.tex
\section{Approach}
\label{sec:approach}
In this section, we introduce the proposed Dual-Graph Sequential Recommender (DGSR) method. We first give the problem formulation. Then, the basic sequence-aware factorization framework of DGSR is introduced. After that, we explain how to build two types of graph and incorporate them for user and item embedding enhancement. Finally, we introduce the prediction and optimization of the proposed method.

\subsection{Problem Formulation}
The problem studied in this paper is sequential fashion recommendation. Let $\mathcal{U}=\{u_1, u_2, ...u_{N_U}\}$ denote the whole user set and $\mathcal{I}=\{i_1, i_2, ...i_{N_I}\}$ denote the whole fashion item set, where $N_U$ and $N_I$ are the total number of users and fashion items respectively. Each user $u$ is interacted with a sequence of fashion items in chronological order, denoted as $\textbf{s}^u = [s_1^u, s_2^u, ..., s_t^u, ..., s_{|s^u|}^u]$, $s_t^u \in \mathcal{I}$. The objective of sequential fashion recommendation is to predict the next action and make recommendations accordingly, in other words, to predict the probability of item $i$ to be the next pick of $u$ after picking $l$: $y_{u,l,i}:=p(i =s_t^u|l=s_{t-1}^u,u)$.

\subsection{Basic Sequential Recommendation Framework}
Personalized sequential recommendation aims to predict the triplet score among the user, previously interacted item and target item $y_{u,l,i}$. It can be formulated as a personal transition cube prediction problem~\cite{rendle2010factorizing}, where one own transition matrix is learned for each user. 
Due to the limited observations for estimating the transition cube, factorizing methods are adopted to factorize the transition cube into a linear combination of both user-item (u-i) interaction and item-item (i-i) transition~\cite{rendle2010factorizing}: $y_{u,l,i} = y_{u,i} + y_{l,i}$. Following the typical sequential recommendation approach FPMC~\cite{rendle2010factorizing}, we describe a user (or an item) with an embedding vector and model the u-i interaction and i-i transition with inner-product as: 
\begin{equation}
    y_{u,l,i} = \pmb{e}^U_u  \cdot \pmb{e}^{IU}_i+\pmb{e}^L_l \cdot \pmb{e}^{IL}_i.
    \label{eq1}
\end{equation}
$\pmb{e}^U_u \in \mathbb{R}^d$ is the embedding of user $u$, $\pmb{e}^{IU}_i \in \mathbb{R}^d$ is the embedding of item $i$ specifically for the modeling of the interaction with users. $\pmb{e}^U_u$ and $\pmb{e}^{IU}_i$ are looked up from the embedding table of users $\pmb{E}^U \in \mathbb{R}^{N_U \times d}$ and that of items $\pmb{E}^{IU} \in \mathbb{R}^{N_I \times d}$, where $d$, $N_U$ and $N_I$ are the the embedding dimensionality, user set length, and item set length respectively. 
Likewise, $\pmb{e}^L_l \in \mathbb{R}^d$ is the embedding of item $l$, to specifically represent the anchor (last interacted) items, $\pmb{e}^{IL}_i \in \mathbb{R}^d$ is the embedding of item $i$ that is specifically for the item transition modeling. The corresponding embedding tables are $\pmb{E}^L \in \mathbb{R}^{N_I \times d}$ and $\pmb{E}^{IL} \in \mathbb{R}^{N_I \times d}$. Note that although all the $\pmb{E}^{IU}$, $\pmb{E}^L$ and $\pmb{E}^{IL}$ are for item embedding lookup, they are individually initialized and optimized for different purposes.

After the factorization operation, the overall prediction score is divided into two parts: 1) the interaction probability score between the given user and the target item, corresponding to the user's personalized long-term static preference, and 2) the transition probability score between the previously interacted item and the target item, corresponding to the user's short-term dynamic preference. 

Even though FPMC is a concise and effective method, it has the following inherent limitations. First, the user and item representative capability is limited since the model learns the user/item representations solely based on the dot-product score of specific interaction (transition) pair while overlooking the global contextual information (\textit{i.e.}, all interaction or transition pairs). For the u-i interaction, all the items (users) a user (an item) has interacted with can be treated as the features of the it~\cite{wang2019ngcf}. Such features can enhance the user (item) representations. Likewise, for the i-i transition, all the items that an item connected (either as predecessors or successors) can be treated as the contextual features, which can be used to enhance the short-term preference modeling. 

Second, FPMC only considers first-order relations, \textit{i.e.}, the direct u-i interaction and first-order Markov Chain transition between item and item, while overlooking the higher-order relations. In fact, higher-order relations are prevalent and meaningful in modeling both u-i interaction and i-i transition. As shown in Figure~\ref{fig:graph} (we will present the details of the graph construction later), $u_1$ and $u_2$ have a second-order connection since both of them have interacted with item $i_1$. Consequently, $u_1$ and $u_2$ may share some common preferences and could serve as CF features for each other. Similarly, the higher-order Markov Chain transitions also provide rich contextual information to improve the modeling of the i-i transition. Based on above considerations of learning better user and item representations, we propose to leverage two types of global graph to exploit more higher-order relations. 

\vspace{-5pt}
\subsection{User-Item Interaction Graph}
\textbf{Graph Construction: }
We demonstrate a toy example of sequential fashion recommendation in Figure~\ref{fig:graph} (a), where there are two users ($u_1$ and $u_2$) and each of them has a sequence of interacted fashion items. The following (both u-i and i-i) graph construction illustrations are based on this example. To build the u-i interaction graph, we follow the previous GNN-based recommendation approaches~\cite{wang2019ngcf,he2020lightgcn} and build a bipartite graph, as shown in Figure~\ref{fig:graph} (b). Two types of nodes, namely user nodes and item nodes, as well as the interactions (edges) between the users and items form the graph. The first-order connection is built upon the interaction relationship between users and items. Higher-order connectivity can be derived by combining two or more edges consecutively. For example, $u_1$ and $u_2$ can be connected in a second-order relation through their connection with item $i_1$. Such a second-order connection is meaningful as it suggests common choice and preference.
Similarly, items $i_1$ and $i_4$ are also connected in second-order as they are both interacted with $u_2$, which means that the two items may be related in some dimensions, such as with similar style. 

\par \noindent \textbf{Graph Convolution: } 
To enhance the user and item embeddings, we perform information propagation over the u-i interaction graph. We adopt the implementation of  LightGCN~\cite{he2020lightgcn}, which drops the feature transformation, non-linear activation and the self-connection while just keeps the simple weighted sum aggregator. The $k$-th order information propagation is denoted as: 
\begin{equation} \label{eq_2}
\left\{
\begin{aligned}
    \pmb{e}_u^{(k)}=\sum_{i \in \mathcal{N}_u}{\frac{1}{\sqrt{|\mathcal{N}_u|}\sqrt{|\mathcal{N}_i|}}\pmb{e}^{(k-1)}_i}, \\
    \pmb{e}_i^{(k)}=\sum_{u \in \mathcal{N}_i}{\frac{1}{\sqrt{|\mathcal{N}_i|}\sqrt{|\mathcal{N}_u|}}\pmb{e}^{(k-1)}_u},
\end{aligned}
\right.
\end{equation}
where $\mathcal{N}_u$ and $\mathcal{N}_i$ denote the item set connected with user $u$ and the user set connected with item $i$. Note that the $\pmb{e}_i$ here is a simplification of $\pmb{e}_i^{IU}$. The symmetric normalization term $\frac{1}{\sqrt{|\mathcal{N}_u|}\sqrt{|\mathcal{N}_i|}}$ follows the design of standard GCN~\cite{kipf2016gcnsemi}, which prevents the embeddings' scale accumulation after the information propagation~\cite{he2020lightgcn}.

\par \noindent \textbf{Layer Combination: }
We aggregate the information propagated from different order's neighbors by a simple sum operation and obtain the final user and item representations as follows:
\begin{equation}
    \pmb{e}^{U*}_u = \sum^K_{k=0}{\pmb{e}^{(k)}_u};\ \ \ \ \pmb{e}^{IU*}_i = \sum^K_{k=0}{\pmb{e}^{(k)}_i}.
\end{equation}
$\pmb{e}^{(k)}_u$ and $\pmb{e}^{(k)}_i$ are the propagated information from the $k$-th layer of graph convolution for user and item respectively, and $K$ is the number of graph convolution layers (highest connection orders).

\subsection{Item-Item Transition Graph}
\par \noindent \textbf{Graph Construction: }
To build the i-i transition graph, several designs have been proposed before~\cite{xu2019graph,wu2019session,wu2020graph}. For example, Wu \textit{et al.} build a session-level graph for each session to model the higher-order transitions, \textit{i.e.}, SR-GNN~\cite{wu2019session}. However, such session graphs cannot model the cross-session transitions and overlook the global contextual information. In this paper, we propose a new design of i-i transition graph. Similar with the u-i interaction graph, we also formulate the i-i transition graph as a bipartite graph, as shown in Figure~\ref{fig:graph} (c). For each item, we have two types of nodes $l$ and $i$, corresponding to different situations of items for being the anchor or the target respectively. The connections (edges) are the first-order transitions between two adjacent items in the interaction sequence. We perform information propagation from anchor nodes to target nodes, resulting in the enhanced anchor representations by incorporating all of its global successors. Similarly, we also perform information propagation from target nodes to anchor nodes, resulting in the enhanced target representations by incorporating all of its global predecessors. 

Figure~\ref{fig:graph} (c) illustrates the constructed i-i transition graph for the toy example and is associated with the first-order and second-order adjacent matrices. Higher-order connectivity can be derived by combining multiple consecutive connections. For example, $i_1$ and $i_3$ are connected in second-order by the middle node $i_2$ since $[i_1, i_2, i_3]$ is a sequence of $u_i$. Moreover, $i_4$ and $i_2$ also have a second-order connection by the middle node $i_1$ since $i_4$ transits to $i_1$ in the interaction sequence of $u_2$ and $i_1$ transits to $i_2$ in the sequence of $u_1$. This example indicates that our global graph can model cross-session i-i transitions. For simplicity, we just demonstrate the partial transitions from anchor to target, and the transitions from target to anchor are similar. Note that our design not only models the transition directions but also has better representative capability since we further specify the item representation with respect to its different situations.

\par \noindent \textbf{Graph Convolution: } 
Several graph kernels for i-i transitions are utilized by previous methods, such as GGNN in SR-GNN~\cite{wu2019session}. However, existing widely applied kernels, \textit{e.g.,} GGNN ~\cite{wu2019session} or GAT~\cite{qiu2019rethinking}, introduce extra computational costs, which makes it harder to effectively train the models. Moreover, inspired by the LightGCN~\cite{he2020lightgcn}, removing the redundant modules of GCN may boost the recommendation performance. Therefore, in this paper, we adopt the same graph convolution kernel of LightGCN for information propagation over the i-i transition graph. The $k$-th order information propagation is conducted by:
\begin{equation}
\left\{
\begin{aligned}
    \pmb{e}_l^{(k)}=\sum_{i \in \mathcal{N}_l}{\frac{1}{\sqrt{|\mathcal{N}_l|}\sqrt{|\mathcal{N}_i|}}\pmb{e}^{(k-1)}_i}, \\
    \pmb{e}_i^{(k)}=\sum_{l \in \mathcal{N}_i}{\frac{1}{\sqrt{|\mathcal{N}_i|}\sqrt{|\mathcal{N}_l|}}\pmb{e}^{(k-1)}_l},
\end{aligned}
\right.
\end{equation}
where $\mathcal{N}_l$ and $\mathcal{N}_i$ denote the item set that transit from item $l$ and the item set that transit to item $i$ (the $\mathcal{N}_i$ here is different from that in Equation~\ref{eq_2}). Note that the $\pmb{e}_l$ and $\pmb{e}_i$ here are simplifications of $\pmb{e}_l^{L}$ and $\pmb{e}_i^{IL}$ respectively. The symmetric normalization term $\frac{1}{\sqrt{|\mathcal{N}_l|}\sqrt{|\mathcal{N}_i|}}$ follows the design of standard GCN~\cite{kipf2016gcnsemi}.

\par \noindent \textbf{Layer Combination: }
Similar to the u-i graph, to obtain the final anchor item and target item representations for item-item transition, we aggregate the propagated information from neighbors in different orders by a simple sum operations: 
\begin{equation}
    \pmb{e}^{L*}_l = \sum^K_{k=0}{\pmb{e}^{(k)}_l};\ \ \ \ \ \pmb{e}^{IL*}_i = \sum^K_{k=0}{\pmb{e}^{(k)}_i},
\end{equation}
where $\pmb{e}^{(k)}_l$ and $\pmb{e}^{(k)}_i$ are the propagated information from the $k$-th layer of graph convolution for the anchor item and the target item respectively, and $K$ is the number of graph convolution layers.

\subsection{Prediction and Optimization}
After the information propagation over two graphs, we therefore obtain the updated user and item representations. The prediction of the score for the target item $i$ to be the next pick by user $u$ after item $l$ can be computed as:
\begin{equation}
    y^*_{u,l,i} = \pmb{e}^{U*}_u  \cdot \pmb{e}^{IU*}_i+\pmb{e}^{L*}_l \cdot  \pmb{e}^{IL*}_i.
    \label{eqn}
\end{equation}
We apply the pairwise BPR loss~\cite{rendle2012bpr} to train the model, treating all observed triplets as positive samples and unobserved triplets as negatives. For each positive sample $(u,l,i)$, we randomly sample one negative triplet $(u,l,j)$ with same user $u$ and anchor item $l$. The BPR loss is specifically calculated by:
\begin{equation}
    Loss = \sum_{(u,l,i,j) \in \mathcal{O}}{-\text{ln}\sigma(y^*_{u,l,i} - y^*_{u,l,j})} + \lambda \Vert \Theta \rVert,
\end{equation}
where $\mathcal{O} = {(u,l,i,j)|u \in \mathcal{U}, l=s^u_{t-1}, i=\textbf{s}^u_{t}, t \in [0, |\textbf{s}^u|]}, j \notin \textbf{s}^u$. 
$\sigma (\cdot)$ is the sigmoid function; $\Theta$ denotes all trainable parameters in the model, which include user and item embeddings. $\lambda$ is the hyper-parameter controlling the regularization strength. 

%% file: 4_experiments.tex
\section{Experiments}
\label{sec:experiment}
\begin{table}
  \caption{The properties of the chosen methods in terms of (a) P: personalized, (b) S: sequence-aware, (c) $\text{IP}_{ui}$: considering information propagation between user and items, and (d) $\text{IP}_{ii}$: information propagation between items.}
  \vspace{-0.15in}
  \begin{center}
  \label{tab:statistics}
  \begin{tabular}{c p{15mm}p{5mm}p{5mm}p{5mm}p{5mm}}
    \hline
    Type &Model &P &S &$\text{IP}_{\text{ui}}$ ~&$\text{IP}_{\text{ii}}$ \\ 
\hline
\multirowcell{3}{MF-based}& MF &\checkmark & & & \\
& FMC & &\checkmark & & \\
& FPMC &\checkmark &\checkmark & & \\
\hline
\multirowcell{2}{Non-sequential \\with GNN}& NGCF &\checkmark & &\checkmark &  \\
& LightGCN &\checkmark & &\checkmark  & \\
\hline
\multirowcell{3}{Session-based\\ with DNN}& GRU4REC &\checkmark &\checkmark &  & \\
& Caser &\checkmark &\checkmark & &  \\

& SR-GNN& &\checkmark & &\checkmark \\
\hline
Ours &DGSR  &\checkmark &\checkmark &\checkmark &\checkmark \\
    \hline
    \label{tab:baselines}
  \end{tabular}
  \vspace{-0.2in}
  \end{center}
\end{table}
In this section, we conduct experiments on two real-world fashion e-commerce datasets to evaluate the effectiveness of the proposed method in sequential fashion recommendation. We particularly focus on the following three research questions:
\begin{itemize}
    \item \textbf{RQ1: } Does the proposed DGSR method outperform state-of-the-art sequential and non-sequential fashion recommendation methods? 
    \item \textbf{RQ2: } Does each technical component in the model work, and how does the hyper-parameters and specific settings affect the performance of the model? 
    \item \textbf{RQ3: } How does the proposed method perform on data with different sparsity, and how does the incorporated graphs affect the recommendation results? 
\end{itemize}

\begin{table*}
\caption{Overall recommendation performance}
\vspace{-0.1in}
\begin{center}
\setlength{\tabcolsep}{2.1mm}{\begin{tabular}{{cccc|ccc||ccc|ccc}}
\hline
~ &\multicolumn{3}{c|}{iFashion-SR-7} &\multicolumn{3}{c||}{iFashion-SR-10} &\multicolumn{3}{c|}{Amazon-Fashion-7} &\multicolumn{3}{c}{Amazon-Fashion-10}\\
Model &Recall &MRR &NDCG &Recall &MRR &NDCG &Recall &MRR &NDCG &Recall &MRR &NDCG \\
\hline
MF &0.5929 &0.4421 &0.4778 &0.5938 &0.4434 &0.4791 &0.3382 &0.2044 &0.2357 &0.2606 &0.1444 &0.1716 \\
FMC &0.4691 &0.3316 &0.3641 &0.4735 &0.3338 &0.3673 &0.2362 &0.1376 &0.1606 &0.1534 &0.0764 &0.0942\\
FPMC &0.5894 &0.4421 &0.4771 &0.6006 &0.4372 &0.4762 &0.3436 &0.2078 &0.2397 &0.2643 &0.1513 &0.1777\\
GRU4REC &0.5149 &0.3526 &0.3911 &0.4942 &0.3402 &0.3768 &0.3663 &0.2301 &0.2622 &0.2599 &0.1317 &0.1615\\
Caser &0.4410 &0.3179 &0.3471 &0.4275 &0.3012 &0.3311 &0.4000 &0.2338 &0.2730 &0.2772 &0.1633 &0.1898\\

LightGCN &0.5928 &0.4414 &0.4775 &0.5767 &0.4240 &0.4580 &0.3803 &0.2373 &0.2769 &0.3097 &0.1784 &0.2092\\
NGCF &0.5897 &0.4186 &0.4593 &0.5923 &0.4182 &0.4595 &\textbf{0.4081} &\textbf{0.2634} &\textbf{0.2974} &\textbf{0.3334} &\textbf{0.1982} &\textbf{0.2299}\\
SR-GNN &0.5485 &0.3862 &0.4248 &0.5494 &0.3890 &0.4271 &0.3900 &0.2566 &0.2884 &0.3125 &0.1956 &0.2224\\
\text{DGSR} &\textbf{0.6478} &\textbf{0.5027} &\textbf{0.5373} &\textbf{0.6522} &\textbf{0.5045} &\textbf{0.5398} &0.3851 &0.2390 &0.2733 &0.3125 &0.1822 &0.2127\\

\hline
\end{tabular}}
\label{tab:overall-performance-ifashion}
\end{center}
\end{table*}

\subsection{Dataset}
We evaluate our proposed model based on two real-world e-commerce datasets: iFashion~\cite{chen2019pog} and Amazon review datasets~\cite{mcauley2015image}. 

\textbf{iFashion} is a large-scale fashion dataset generated based on the e-commerce platform Taobao\footnote{www.taobao.com}. The original dataset contains over three million users and four million items, where each user is interacted with tens of items in chronological order. To make the datasets applicable to all sequential recommendation methods, we crop two sub-datasets which consist of different minimal fixed-length user interaction sequences of 7 and 10 respectively. For each sub-dataset, we randomly sample 50,000 sequences from the interaction sequence list, whose length must be longer than the predefined minimal sequence length. Since one user may have multiple interaction sequences in the dataset, the user number is less than the sequence number. The generated dataset is named as \textbf{iFashion-SR-7} and \textbf{iFashion-SR-10} respectively, where iFashion-SR denotes \textbf{iFashion} for \textbf{S}equential \textbf{R}ecommendation.

\textbf{Amazon review dataset} is a large-scale review dataset which consists of the purchase records and reviews of users on Amazon\footnote{www.amazon.com}. We use the \textit{Clothing-Shoes-Jewelry subset data (2014 version)} to generate our dataset for sequential recommendation. Since the long-tail situation is severe in this dataset, we filter out the items with less than five interactions at the very first step. Similar with the preprocessing for iFashion dataset, we generate two sub-datasets for users with interacted sequence length longer than the minimal predefined length of 7 and 10 respectively. 

For each data setting, we split the interaction sequences into three parts: (1) the last interaction in each sequence is used for testing; (2) the second to last interaction in each sequence is used for validation; and (3) all remaining interactions are used for training. The statistics of the datasets are shown in Table~\ref{tab:dataset}. For methods modeling one-step transitions (such as FPMC), the previous one item of each interaction is employed to build a pair. For methods need to model the whole historical sequence before the target item (such as SR-GNN), we generate short sub-sequences with the sliding window strategy~\cite{tang2018personalized} from the original sequences.

\subsection{Experimental settings}
\textbf{Baselines:} We compare the proposed methods with several competitive and relevant baselines:
\begin{itemize}
    \item \textbf{MF}~\cite{rendle2012bpr} factorizes the u-i interaction matrix and is optimized with the BPR loss. 
    \item \textbf{FMC}~\cite{rendle2010factorizing} focuses on modeling the sequential dynamics by factorizing the i-i transition matrix.
    \item \textbf{FPMC}~\cite{rendle2010factorizing} models both the personalized u-i interaction and the `global' i-i transition by MF and FMC respectively.
    \item \textbf{NGCF}~\cite{wang2019ngcf} leverages a bipartite graph to model the u-i interaction, which is a typical GCN-based recommendation model.
    
    \item \textbf{LightGCN}~\cite{he2020lightgcn} is the light version of NGCF and utilizes a light graph convolution kernel. 
    \item \textbf{GRU4REC}~\cite{hidasi2015session} is an RNN-based session-based recommendation method which models the interaction sequences with GRUs. 
    \item \textbf{Caser}~\cite{tang2018personalized} is a session-based recommendation method which treats the embedding of fixed length of historical interacted items as an ``image'' and models the interaction sequences with CNNs.
    \item \textbf{SR-GNN}~\cite{wu2019session} is a GNN-based method which leverages the session-level i-i transition graph to enhance the session representation.
\end{itemize}
Table 2 lists the properties of these methods in terms of whether being personalized or sequence-aware and whether considering the user-item or the item-item information propagation. It offers an intuitive illustration of the difference of those methods.

\noindent{\textbf{Implementation details: }} 
We perform grid search on the embedding size of users and items, which is within [5, 10, 20, 50] . In terms of the number of propagation layers of the u-i graph and i-i graph, we perform grid search within [1, 2, 3] for each type of graph respectively. The learning rate during training is set to 0.01 and the batch size is set to 5000 for all methods. The weight decay is set to $10^{-5}$ and the maximum number of training epoches is 250. We implement all the baseline methods on these two datasets, and for fair comparison, we also do grid search for the user and item embeddings within [5, 10, 20, 50] to achieve the best performance. The hidden size of GRU4REC is chosen from [10, 20, 50, 100]; the number of vertical filters and horizontal filters of Caser are set to 4 and 8 respectively.

\noindent\textbf{Evaluation metrics: }We evaluate the performance of all compared recommendation methods with the mainstream top-$n$ ranking evaluation. Considering the computational cost during the evaluation, we generate a negative set with 100 negative samples for each testing samples (positive) and rank the positive sample along with negative samples based on the predicted scores. All negative items are not interacted with the corresponding user and the anchor item. We apply three commonly used metrics for the quantitative evaluation: Recall@$n$, MRR@$n$, and NDCG@$n$. We set $n=10$. 
\begin{table*}[!t]
\caption{Comparison between models with/without two types of graph. Embedding size is set to 50.}
\vspace{-0.1in}
\begin{center}
\setlength{\tabcolsep}{2.1mm}{\begin{tabular}{{cccc|ccc||ccc|ccc}}
\hline
~& \multicolumn{3}{c|}{iFashion-SR-7} &\multicolumn{3}{c||}{iFashion-SR-10} &\multicolumn{3}{c|}{Amazon-Fashion-7} &\multicolumn{3}{c}{Amazon-Fashion-10}\\

Model &Recall &MRR &NDCG &Recall &MRR &NDCG &Recall &MRR &NDCG &Recall  &MRR &NDCG \\
\hline
\text{DGSR-UI/II} &0.5894 &0.4421 &0.4771 &0.6006 &0.4372 &0.4762 &0.3436 &0.2078 &0.2397 &0.2643 &0.1513 &0.1777\\
\text{DGSR-II} &0.6146 &0.4734 &0.5070 &0.6165 &0.4757 &0.5092 &0.2984 &0.1652 &0.1963 &0.2350 &0.1218 &0.1481\\  
\text{DGSR-UI} &0.6339 &0.4583 &0.5207 &0.6380 &0.4891 &0.5247 &0.3525 &0.2169 &0.2487 &0.2772 &0.1551 &0.1837\\ 
\text{DGSR} &\textbf{0.6478} &\textbf{0.5027} &\textbf{0.5373} &\textbf{0.6522} &\textbf{0.5045} &\textbf{0.5398}  &\textbf{0.3851} &\textbf{0.2390} &\textbf{0.2733} &\textbf{0.3125} &\textbf{0.1822} &\textbf{0.2127}\\ 

\hline
\end{tabular}}
\label{tab:ablation-graph}
\end{center}
\end{table*}

\subsection{Overall performance (RQ1)}
Table~\ref{tab:overall-performance-ifashion} shows the overall performance of all compared methods on the iFashion-SR and Amazon-Fashion datasets. We have the following observation based on the results.

\noindent \textbf{Performance on iFashion-SR: } 
(1) Our method outperforms all the compared methods in both settings in terms of all three metrics by a large margin.

(2) Compared to the basic FPMC, which also models both user-item interaction and item-item transition, our method achieves around 10\% improvement in terms of NDCG for both settings due to the incorporation of two types of global graph. 

(3) LightGCN shows the best performance among all baselines on both settings, while the simple MF and FPMC are competitive too according to the results. Such results may imply that the collaborative signals (CF) in this dataset are quite significant. It also shows that modeling u-i interaction is effective on this dataset. 

(4) Session-based methods, including GRU4REC, Caser and SR-GNN, do not show superior performance as expected. The reasons are multi-fold. First, most of them (except Caser) ignore user information in the model. Second, the training samples are less for them as they require to input multiple historical interactions. Third, they ignore the effect among the connected users and items in higher orders.

\noindent \textbf{Performance on Amazon-Fashion: } 
(1) DGSR shows competitive performance on both settings and outperforms most CF-based methods, demonstrating the effectiveness of it to some extent. 

(2) Overall, session-based methods perform better than methods such as FMC, which suggests that the i-i transition signals are harder to explore for this dataset. Since session-based methods are better in exploring the dependency among items by considering the item transition sequences rather than the pairs in the model, they achieve more desired performance.

(3) As the i-i transition signals are less effective to model than the u-i interaction, our DGSR, exploring more high-order i-i transitions, might affect the exploring o u-i patterns instead, which degrades the overall performance of DGSR.





\begin{figure}
    \centering
    \includegraphics[width = 0.9\linewidth]{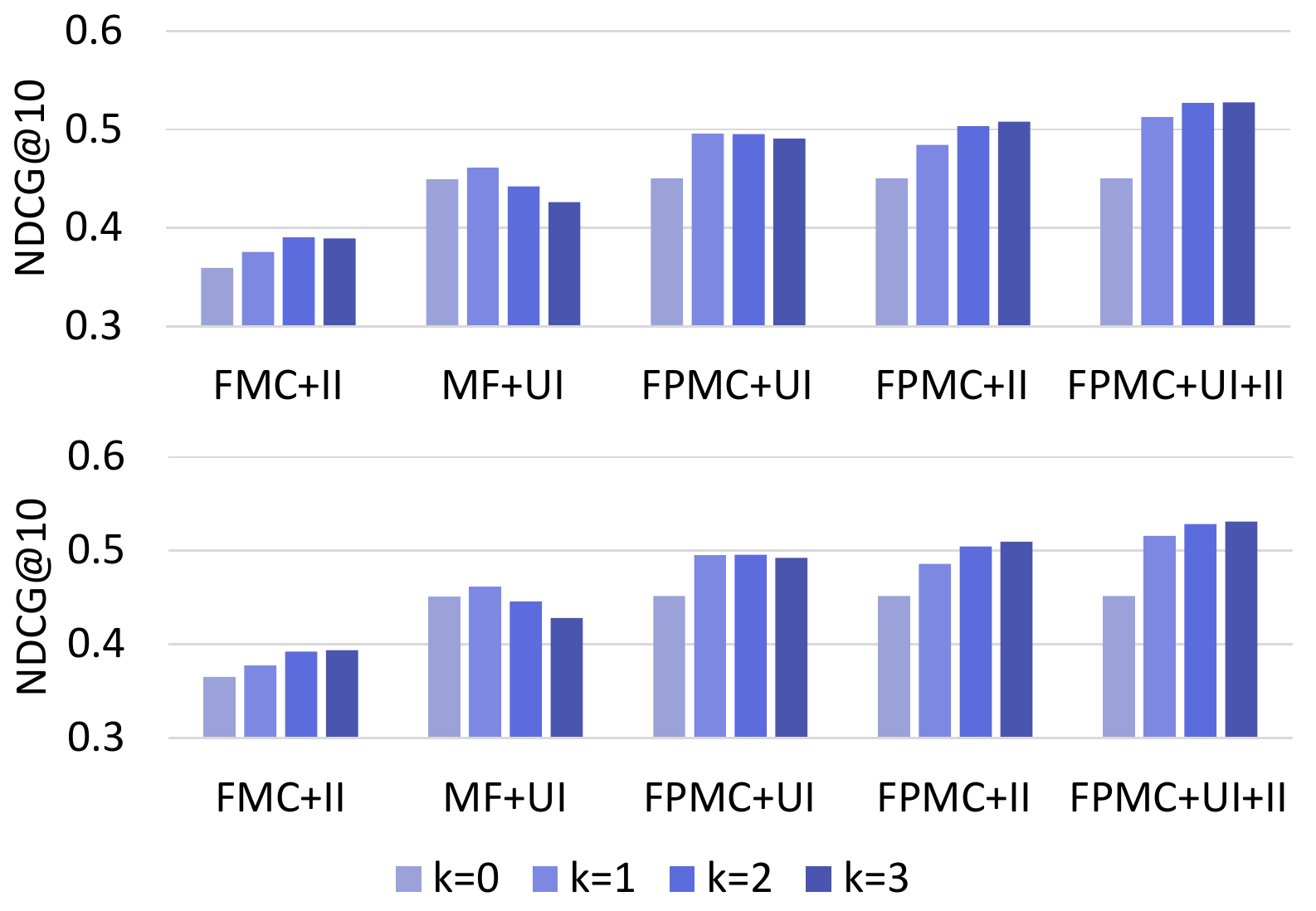}
    \caption{Performance of models with various graph layers. Up: iFashion-SR-7; Down: iFashion-SR-10. $k$ denotes the number of the layers for information propagation. FPMC+UI+II is the DGSR method.}
    \label{fig:layers}
\end{figure}

\subsection{Ablation study (RQ2)}
In this part, we conduct several ablation studies to discuss more detailed technical components in DGSR. The ablation study focuses on three aspects: 1) the effect of incorporating two types of graph; 2) the effect of the layer numbers ($k$) in each graph; and 3) the effect of embedding size.

\noindent \textbf{Effect of global graphs:}
We first investigate the effectiveness of the two types of graph, namely, the u-i interaction graph and the i-i transition graph. We conduct the comparison on the variant of DGSR in which one or both of the graph are removed. The experimental results on four data settings are listed in Table~\ref{tab:ablation-graph}. 

From the results we can observe that: First, either incorporating u-i graph or i-i graph can improve the performance on iFashion-SR dataset. On Amazon-Fashion, the u-i graph seems not helpful from the comparison between the models with/without u-i graph (the first two rows), while the i-i graph brings significant improvement (comparing first and third rows). As discussed in the above section, modeling i-i transition is hard on Amazon-Fashion, which may even degrade the performance when combining with the u-i interaction modeling (comparing results of MF, FMC and FPMC in Table~\ref{tab:overall-performance-ifashion}). When we further strengthen the u-i modeling, it leads to more conflicts between u-i and i-i modeling. However, when we apply both graphs, the performance of the method on all settings boosts; this is obvious on Amazon-Fashion that the i-i transition modeling significantly improve the overall recommendation performance. 

\begin{figure}
    \centering
    \includegraphics[width = 1\linewidth]{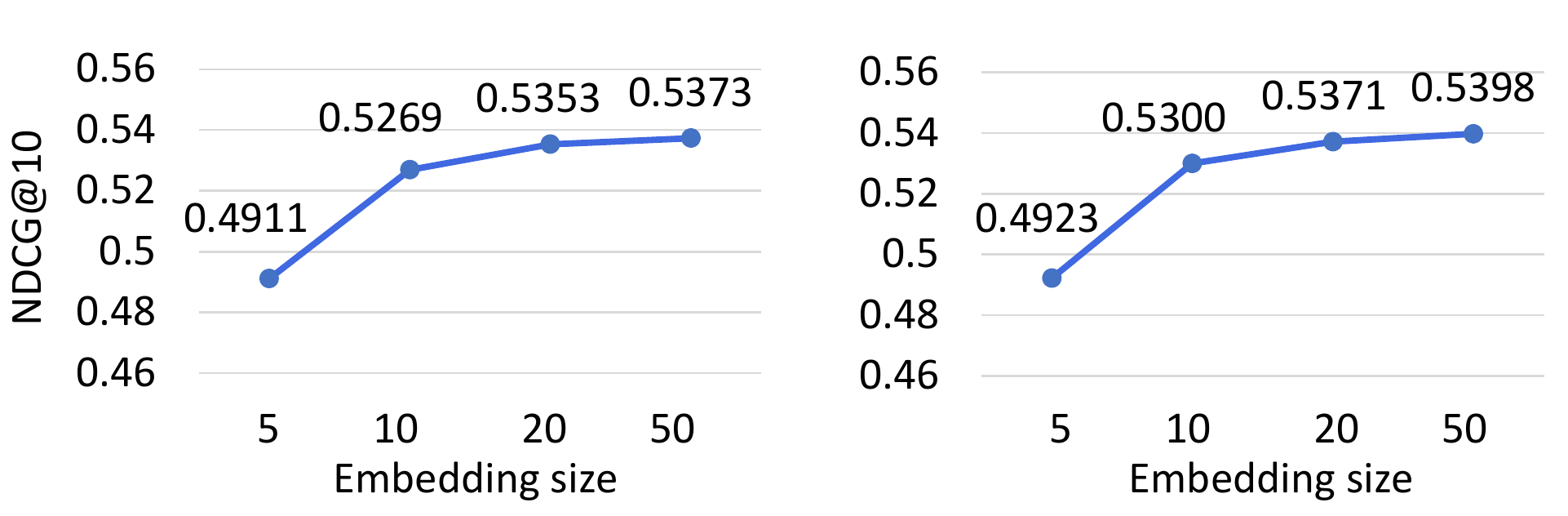}
    \vspace{-0.15in}
    \caption{Performance of DGSR with different embedding sizes. Left: iFashion-SR-7; Right: iFashion-SR-10. }
    \vspace{-0.15in}
    \label{fig:emb}
\end{figure}
\noindent \textbf{Number of propagation layers:}
We further investigate the effect of the number of propagation layers when the two types of graph are applied in various models. We study three basic models, MF which only explores u-i interaction, FMC which only exolores i-i transition, and FPMC which explores both. We apply u-i graph and i-i graph on three models and generate five improved models with graph in total. We set different graph layers for each model from 0 (without graph) to 3, resulting in 15 experimental settings for each dataset. The NGCG@10 results of all compared models on the iFashion-SR dataset are shown in Figure~\ref{fig:layers}.

From the results we have the following observations. 1) The performance of all three basic models are improved by leveraging various graphs. The best layer number setting for different models are different. 2) For the FMC model, the performance increases with the increase of graph layers. This demonstrates that higher-order connectivity is important to improve the i-i transition pattern modeling. 3) For MF+UI and FPMC+UI, the best performance is achieved when one-layer information propagation is conducted, which shows that stacking more graph layers may over-smooth the representation of users and items and degrade the performance instead. 4) When leveraging both graphs, we can see that the performance steadily improves with the increase in layer number, which shows that the u-i graph and i-i graph organically corporate with each other and contribute to the final interaction prediction modeling.

\begin{figure}
    \centering
    \includegraphics[width = 1.0\linewidth]{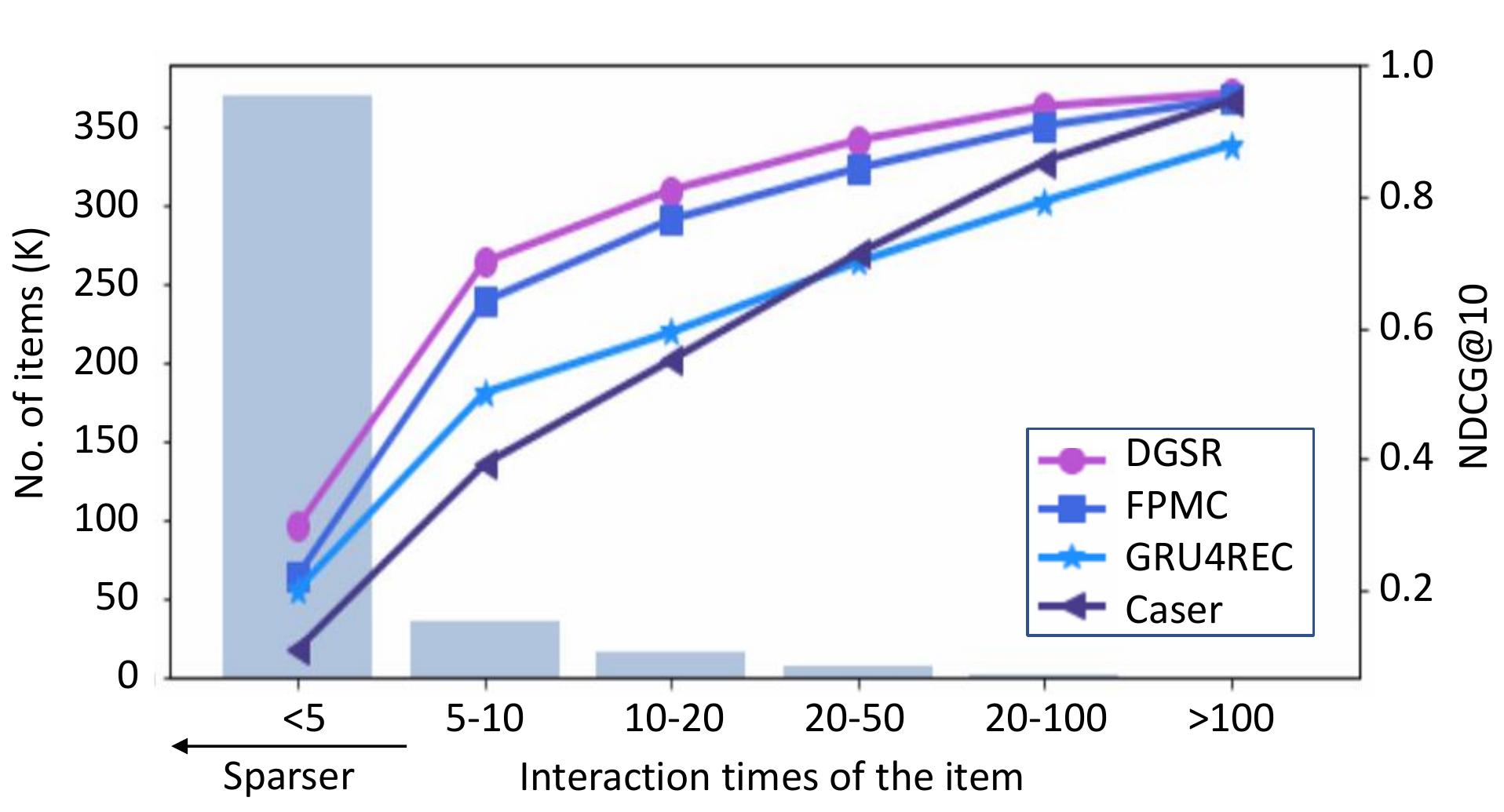}
    \vspace{-0.2in}
    \caption{Performance comparison w.r.t interaction sparsity. The comparison is based on the NDCG results on iFashion-SR-7. The distribution of the items in terms of the interaction density is also shown by the bar chart in the background. }
    \vspace{-0.15in}
    \label{fig:density}
\end{figure}

\noindent \textbf{Effect of embedding size: }
In Figure~\ref{fig:emb}, we illustrate the performance of the proposed model with different embedding sizes. From the results we can see that larger embedding size does bring increase in performance. However, the rate of performance boost keeps decreasing when the embedding size gets larger. In the meantime, using large embedding size inevitably results in large computational cost, which reduces the efficiency of the model. Therefore, we think the embedding size of 10 is quite an ideal setting when considering both the effectiveness and efficiency.

\subsection{Discussion on Sparsity and Higher-order Connectivity (RQ3)}
In this section, we first discuss the sparsity problem by showing the performance of DGSR and other three baselines on testing samples with different levels of sparsity. From Figure~\ref{fig:density}, we can see that generally, all methods perform better on dense samples and worse on sparse ones, which is consistent with our hypothesis. The DGSR method outperforms the others in the whole density range, but the superiority is more significant on sparse data (items interacted lower than 10). Such a result demonstrates that our proposed DGSR method can effectively tackle the data sparsity problem as claimed.

\begin{figure}
    \centering
    \includegraphics[width = 1.0\linewidth]{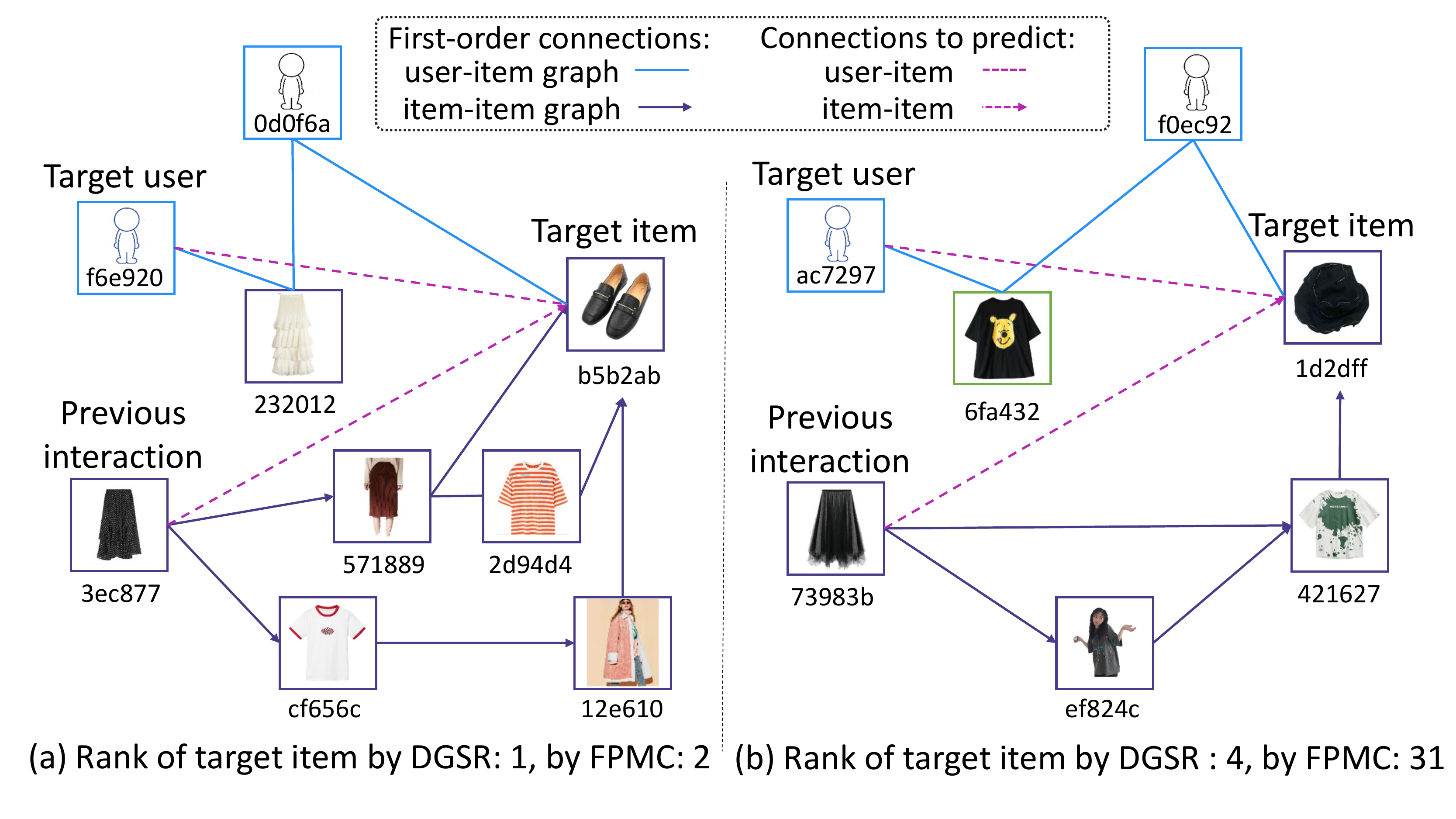}
    \vspace{-0.2in}
    \caption{Examples of test samples and the connectivity in two graphs.}
    \vspace{-0.1in}
    \label{fig:case}
\end{figure}

We further analyze the higher-order connectivity in the data and the specific effect of it on the prediction and recommendation. We illustrate two cases in Figure~\ref{fig:case} and show the \textbf{target user}, the \textbf{previously interacted item}, the \textbf{target item}, as well as the multiple connectivities related to the three subjects in each case. From the examples, we can see that our model is able to rank the target item higher in both cases with the two graphs. By analyzing more details, we find that the target items in both cases are connected with other user and items by interaction or transition relationships. In the first case, the target item \textit{b5b2ab} is connected to another user \textit{0d0f6a} with the u-i graph, who picks item \textit{232012}. As item \textit{232012} is also picked by the target user \textit{f6e920}, the target user and target item are connected in third-order. With the message passing and aggregation operations, the representation of the target user and item will be closer, so that the final interaction probability predicted by the model will be higher. The higher-order connectivity in i-i graph works in a similar way, which can help improve the prediction of the transition probability from the anchor item to the target item. In general, we can see that with two graphs, the target item in two cases can be ranked higher.

%% file: 5_conclusion.tex
\section{Conclusion}
\label{conclusion}
We worked on the problem of sequential fashion recommendation in this paper, aiming to model both the interaction between user-item and the transition between items. To tackle the specific challenges of data sparsity and make the model simple to learn, we proposed a DGSR model. DGSR leverages two types of graphs, namely the user-item interaction graph and the item-item transition graph, to better model the CF signals and the item transitional patterns and enhance the user and item embeddings. Extensive experiments on two datasets demonstrated the effectiveness of the proposed method and all technical contributions. 

In future work, we shall improve this work in the following directions. First, we plan to incorporate additional information besides the implicit feedback to enrich the graph building. By considering more content information such as attributes or visual features, the connectivity between the user and items or item and items can be richer and thus further help to alleviate the interaction sparsity issue. Second, more fashion domain knowledge can be introduced in the model to further enhance the user and item representation learning to achieve better recommendation performance.

%% file: 6_acknowledgement.tex
\section*{acknowledgement}
This research is supported in part by the National Research Foundation, Singapore under its International Research Centres in Singapore Funding Initiative. Any opinions, findings and conclusions or recommendations expressed in this material are those of the author(s) and do not reflect the views of National Research Foundation, Singapore. It is also supported in part by the Natural Science Foundation of China under Grant 61703283, in part by the Guangdong Natural Science Foundation under Project 2021A1515011318, in part by the Shenzhen Municipal Science and Technology Innovation Council under the Grant JCYJ20190808113411274.